%% file: main.tex
\documentclass[11pt]{article}

    \usepackage[usenames,dvipsnames]{xcolor}
    \usepackage[margin=1in]{geometry}
    \usepackage{graphicx} 
    \usepackage{bm} 
    \usepackage{color} 
    \usepackage{dcolumn}
    \usepackage{siunitx}
    \usepackage[mathlines]{lineno}
    \usepackage{setspace} 
    \usepackage{booktabs}
    \usepackage{subfiles}
    \usepackage{enumitem}
    \usepackage{tikz}
    \usepackage{scalerel}
    \usepackage[version=4]{mhchem}
    \usetikzlibrary{svg.path}
    \usepackage{placeins}
    \usepackage{xspace}
    \usepackage[
        backend=biber,
        style=nature,
        sorting=none,
        giveninits=true,
        maxbibnames=99,
        doi=true,
        url=false,
        isbn=false,
        eprint=false
    ]{biblatex}
    \usepackage[linktocpage,colorlinks=true,linkcolor=blue,citecolor=blue,breaklinks=true,urlcolor=blue]{hyperref}

    \usepackage{caption}
    \newcommand{\ecoli}{\textit{E.~coli}\xspace}
    \newcommand{\vcholerae}{\textit{V.~cholerae}\xspace}
    \newcommand{\paeruginosa}{\textit{P.~aeruginosa}\xspace}
    \newcommand{\senterica}{\textit{S.~enterica}\xspace}

    \DeclareSIUnit{\revolutionsperminute}{rpm}

    \definecolor{orcidlogocol}{HTML}{A6CE39}
    \tikzset{
      orcidlogo/.pic={
        \fill[orcidlogocol] svg{M256,128c0,70.7-57.3,128-128,128C57.3,256,0,198.7,0,128C0,57.3,57.3,0,128,0C198.7,0,256,57.3,256,128z};
        \fill[white] svg{M86.3,186.2H70.9V79.1h15.4v48.4V186.2z}
                     svg{M108.9,79.1h41.6c39.6,0,57,28.3,57,53.6c0,27.5-21.5,53.6-56.8,53.6h-41.8V79.1z M124.3,172.4h24.5c34.9,0,42.9-26.5,42.9-39.7c0-21.5-13.7-39.7-43.7-39.7h-23.7V172.4z}
                     svg{M88.7,56.8c0,5.5-4.5,10.1-10.1,10.1c-5.6,0-10.1-4.6-10.1-10.1c0-5.6,4.5-10.1,10.1-10.1C84.2,46.7,88.7,51.3,88.7,56.8z};
      }
    }
    \newcommand\orcid[1]{\href{https://orcid.org/#1}{\mbox{\scalerel*{
    \begin{tikzpicture}[yscale=-1,transform shape]
    \pic{orcidlogo};
    \end{tikzpicture}
    }{|}}}}

\begin{document}


    \title{Multiflagellarity facilitates bacterial upstream motility}
    \author{%
        \parbox{0.94\textwidth}{%
        \centering
        Ran Tao\textsuperscript{1,\textdagger}\,\orcid{0009-0001-7720-6459},
        Nathaniel C. Esteves\textsuperscript{2,\textdagger},
        Wanho Lee\textsuperscript{3, \textdagger},
        Lauren Altman\textsuperscript{1},\\
        Liuni Chen\textsuperscript{4},
        David Gao\textsuperscript{1,5},
        Ling Li\textsuperscript{4},
        Yongsam Kim\textsuperscript{6},\\
        Jun Zhu\textsuperscript{2,*},
        Sookkyung Lim\textsuperscript{7,*}, and
        Arnold J. T. M. Mathijssen\textsuperscript{1,*}\,\orcid{0000-0002-9577-8928}\\[0.75em]
        \small \textsuperscript{1}Department of Physics \& Astronomy,
        University of Pennsylvania, Philadelphia, PA 19104, USA\\
        \small \textsuperscript{2}Perelman School of Medicine,
        University of Pennsylvania, Philadelphia, PA 19104, USA\\
        \small \textsuperscript{3}National Institute for Mathematical Sciences,
        Daejeon 34047, Republic of Korea\\
        \small \textsuperscript{4}Department of Materials Science and Engineering,
        University of Pennsylvania, Philadelphia, PA 19104, USA\\
        \small \textsuperscript{5}Department of Biology,
        University of Pennsylvania, Philadelphia, PA 19104, USA\\
        \small \textsuperscript{6}Department of Mathematics, Chung-Ang University,
        Dongjak-gu, Heukseok-dong, Seoul 06974, Republic of Korea\\
        \small \textsuperscript{7}Department of Mathematical Sciences,
        University of Cincinnati, Cincinnati, OH 45221, USA\\[0.5em]
        \small \textsuperscript{\textdagger}These authors contributed equally to this work.\\
        \small \textsuperscript{*}Correspondence:
        \href{mailto:junzhu@pennmedicine.upenn.edu}{junzhu@pennmedicine.upenn.edu};
        \href{mailto:sookkyung.lim@uc.edu}{sookkyung.lim@uc.edu};
        \href{mailto:amaths@upenn.edu}{amaths@upenn.edu}
        }%
    }
    \date{}
    
\maketitle

\begin{abstract}
Upstream swimming drives bacterial spreading and surface colonization. Many pathogens encounter fluid flows as they infect the intestines, lungs, and urinary tract, so how bacteria use their flagella to counter these flows matters for disease and treatment. Yet how morphology and flagellar arrangement govern motility against flow remains unknown.
Here, we investigate the biophysical determinants of rheotaxis by combining microfluidics, directed evolution, genetics, holography, and hydrodynamics simulations. Using upstream swimming competitions, we find that peritrichous \textit{E.~coli} and \textit{S.~enterica} rapidly outcompete monotrichous \textit{P.~aeruginosa} and \textit{V.~cholerae}, accumulating upstream at densities up to five orders of magnitude higher, even though \textit{Vibrio} swims three times as fast. Motility selection experiments show that rheotaxis increases with flagellar number and length, confirmed by overexpressing the master regulator \textit{flhD/C}. Three-dimensional holography and single-cell tracking reveal that multiflagellarity stabilizes surface residence and promotes the weathervane effect that reorients cells upstream, a mechanism further supported by simulations that fully resolve flagellar arrangement and fluid-structure interactions. These results establish multiflagellarity as a key facilitator of upstream navigation, governed by near-wall residence and shear-driven reorientation rather than by swimming speed.

\end{abstract}

\section*{Introduction}

Bacteria often reside in flowing environments, from aquatic porous media and marine sediments to host-associated habitats such as mucosal surfaces, urinary tracts, and medical devices~\cite{Wheeler2019NotPlankton, Padron2023BacteriaFlow, Guasto2012FluidMicroorganisms, Persat2015-rl, Lauga2016BacterialHydrodynamics, Figueroa-Morales2020-mb, Dentz2022-yt, Dehkharghani2023-mw, Conrad2018ConfinedBiofilms, Tao2026-yc, Mathijssen2026BiomedicalFunctionalities}. Interestingly, these shear flows act both as a physical barrier to colonization and as a directional hydrodynamic signal that motile bacteria can exploit~\cite{Rusconi2014-cu, Siryaporn2015-mm, Shuppara2025-xl}: Navigation in flows, known as rheotaxis, can promote access to nutrients, escape from predators, and the ability to colonize areas that other species cannot~\cite{Marcos2012-id, Secchi2020-ml,Drescher2013-xn,Cremer2019-cc,Savorana2025-px}. 
Near solid surfaces, rheotaxis can lead to upstream migration, or positive rheotaxis~\cite{Hill2007-wr,Kaya2012-po,Figueroa-Morales2015-fx,Mathijssen2019-zy,Jing2020-mj,Torres-Maldonado2024-mp,Cao2024-nm,Dey2022-sn, Nakane2023RheotaxisGliding}. 
This upstream motion is not necessarily governed by the propulsion speed, but by the ability to turn against the flow direction, and to remain close enough to boundaries where flows are weaker~\cite{Nash2010Run-and-TumbleSwimming, Costanzo2012TransportFlow, Uspal2015RheotaxisWall, Daddi-Moussa-Ider2020-yk}.

The physical basis of bacterial rheotaxis has been studied most extensively in \ecoli, where near-surface swimming and shear-driven torques give rise to a weathervane effect~\cite{Hill2007-wr,Kaya2012-po,Figueroa-Morales2015-fx,Mathijssen2019-zy}: Like a weathervane in the wind, the bacterial flagella turn in the downstream direction, so the cell body points in the upstream direction. 
The pivot point, about which the bacteria turn, is provided by hydrodynamic friction interactions between the cell body and the surface~\cite{Berke2008HydrodynamicSurfaces, Li2009-jh, Drescher2011-jf, Molaei2014-kf, Sipos2015-yu, Bianchi2017-eb}.
This weathervane effect has been visualized directly~\cite{Cao2024-nm} by labeling the flagella fluorescently~\cite{Turner2000-fo, Turner2012GrowthLength}.

However, it remains unclear how upstream swimming is governed by cell morphology and flagellar arrangement, which vary widely across species~\cite{Tokarova2021-wn, Nakane2022-sm, Zhong2025-et, Yang2016StayingEnvironments}. 
It is well-known, at least in the absence of flows, that bacteria strictly regulate their cell body shape and their spatial patterning of flagella and pili ~\cite{Yang2016StayingEnvironments, Kamdar2023-eb, Tian2022-xt, Utada2014-zu, Mears2014-ua, Nguyen2018-ah, Tao2025-er, Lisevich2025-lw}, which directly impact their motility characteristics, including swimming speed and run-tumble behavior~\cite{vanTeeseling2017DeterminantsTargeting, Guadayol2017-od, Grognot2021MoreBehaviors}.

We expect that these features could alter rheotaxis by regulating the weathervane effect, and by tuning the interactions that bacteria have with surfaces and flow gradients.
Yet, systematic comparisons across species and phenotypes have been lacking, and a mechanistic understanding of rheotaxis has been elusive, until now.

\section*{Results}

\subsection*{Multiflagellarity prevails in population-scale upstream swimming competitions}

Bacterial species differ widely in traits we expect to shape how they respond to flow. Therefore, we considered three motile species with distinct morphologies and flagellar arrangements: \ecoli, \paeruginosa, and \vcholerae (Fig.~\ref{fig:1}A-C). Peritrichous \ecoli K-12 cells carry multiple flagella distributed around a rod-shaped cell body and do not express type IV pili, whereas \paeruginosa and \vcholerae cells swim with a single polar flagellum and do express pili ~\cite{Kamdar2023-eb, Tian2022-xt, Utada2014-zu}. The latter also has a curved cell body~\cite{Bartlett2017-vm, Martin2021-oc}. Baseline measurements in flow-free motility chambers showed that all species had comparable cell body dimensions, but differed significantly in shape and intrinsic swimming speed $V_\textmd{swim}$, with \vcholerae moving more than two times faster than \ecoli and \paeruginosa, without flows (Supplementary Fig.~\num{1}).

To compare rheotaxis across these bacterial species under identical flow conditions, we developed a microfluidic upstream swimming competition assay with parallel race tracks (Fig.~\ref{fig:1}D, E, Methods).
Each track consists of a downstream reservoir connected to a series of seventeen cavities, similar to the setup in~\cite{Tao2026-yc}.
Initially, different strains of bacteria were introduced into separate reservoirs and kept there with relatively strong flow.
At the start of the competition, the flow rate was reduced, causing the cavities to fill with cells migrating upstream.
Using fluorescence microscopy, we then quantified the density of bacteria in each cavity over time to monitor their population-level migration throughout the race. 

The outcome of the race is shown in Fig.~\ref{fig:1}D-F.
Interestingly, \vcholerae (yellow) did not move upstream efficiently despite its much higher speed, nor did \paeruginosa (red), while \ecoli (green) progressively moved against the flow and accumulated in the cavities successively.
Within minutes, the cell density of \ecoli in the last cavity (\#17) was a thousand times higher than the other species, and this factor increased to about $10^5$ within two hours (Supplementary Fig.~\num{2}). 

To verify that these behaviors do not depend on the microfluidic channel geometry, we repeated experiments in devices without intermediate cavities and with different channel widths ranging from \SI{10}{} to \SI{50}{\micro\meter} (Supplementary Fig.~\num{3}).
Under all conditions, \ecoli consistently migrated into the upstream reservoir, while \paeruginosa and \vcholerae showed much less upstream motility.
Long-term measurements further showed that \ecoli maintained their accumulation upstream for at least 8 hours, whereas \paeruginosa and \vcholerae showed little detectable accumulation over the same period (Supplementary Fig.~\num{4}).

Together, these assays reveal that upstream migration can drive enormous differences in cell density between species, with \ecoli outnumbering the others by five orders of magnitude.
Yet, these differences cannot be explained by their swimming speed alone. 

\subsection*{Single-cell tracking reveals individual rheotaxis behaviors}

To understand how these large population-level differences arise, we tracked thousands of individual bacteria for each species over long periods of time to compare their swimming behaviors under flow (Fig.~\ref{fig:1}G and Supplementary Fig.~\num{5}).
For this experiment, we used narrow channels (height $H=\SI{10}{\micro\meter}$) to ensure that the bacteria always remained in focus, and without cavities (constant width $W=\SI{25}{\micro\meter}$).

Fig.~\ref{fig:1}H shows representative trajectories for the three species.  
As expected, \ecoli (green) swam upstream very effectively, hopping between the top and bottom edges.
\paeruginosa (red) typically moved downstream, however, with only a few cells going up.
\vcholerae (yellow) performed intermediately, displaying occasional bursts of upstream motion. 

To quantify these dynamics systematically, we measured the average upstream swimming velocity, $\langle V_x \rangle$, for a wide range of shear rates (Fig.~\ref{fig:1}I). 
In weak shear, $\dot{\gamma}\sim \SI{20}{\per\second}$, almost all \ecoli cells moved upstream with $\langle V_x \rangle  \sim V_\textmd{swim}$,
while \vcholerae had a slightly positive $\langle V_x \rangle$ and \paeruginosa slightly negative.
As the imposed shear rate increased, the cells progressively moved more downstream, although \ecoli retained a pronounced upstream bias until high shear rates, around $\dot{\gamma}\sim \SI{80}{\per\second}$.

However, the histograms in Fig.~\ref{fig:1}J show that even for the monotrichous species, the fraction of bacteria moving upstream can be larger than expected:
Because these distributions were highly skewed, 19\% of \paeruginosa bacteria still moved against the flow at $\dot{\gamma} = \SI{53}{\per\second}$, despite a negative mean.
To verify that these results do not depend on confinement, we also tracked bacteria in wider channels, with $H=\SI{50}{\micro\meter}$ and $W=\SI{500}{\micro\meter}$ (Supplementary Fig.~\num{6}). Again, the fraction of \ecoli cells moving upstream was significantly larger than \paeruginosa and \vcholerae.

Overall, these single-cell tracking measurements showed that multiflagellated bacteria can move against faster flows than uniflagellated species.
Yet, even when a population moved downstream on average, individual cells could still persevere.
These observations indicate that upstream performance depends on how cell morphology and flagellar arrangement are coupled with boundary-mediated alignment against shear flow.

\subsection*{Directed evolution with \ecoli shows that rheotaxis is enhanced by hyperflagellation}

While the previous results demonstrated large differences in rheotaxis between peri- and monotrichous bacteria, they could not isolate the effects of flagellar abundance and other species-specific traits.
Therefore, we performed directed evolution assays~\cite{Liu2019-gj} with \ecoli to test whether upstream swimming could be tuned within a single species (Fig.~\ref{fig:2}A).
In these experiments, wild-type cells were inoculated in the center of Petri dishes filled with semi-solid agar, known as swim plates. 
After 24 hours, the rapidly and slowly expanding cells were selected from the outer and inner regions, respectively, and reinoculated onto fresh plates. 
This selection process was repeated for 30 days, after which we isolated the most and least motile strains, named phenotype F and phenotype A, abbreviated $\Phi$F and $\Phi$A, respectively (Supplementary Fig.~\num{7}). 

Following this motility selection, we first verified that $\Phi$F expanded more rapidly than wild-type cells in semi-solid agar, whereas $\Phi$A expanded more slowly (Fig.~\ref{fig:2}B and Supplementary Fig.~\num{7}). We also measured their swimming speed in aqueous BMB using single-cell tracking in flow-free motility chambers (Fig.~\ref{fig:2}C).
$\Phi$F was indeed slightly faster, but the difference was markedly smaller than that observed in semi-solid agar. 

Crucially, we then asked whether these selected strains differed in their rheotactic responses, both at the single-cell and population levels. 
Tracking individual bacteria (Fig.~\ref{fig:2}D,E) revealed that $\Phi$F exhibited the highest upstream swimming velocity, wild-type cells showed intermediate behavior, and $\Phi$A showed reduced rheotaxis. This hierarchy was maintained across different shear rates and channel widths (Supplementary Figs.~\num{8} and \num{9}, respectively). 
At the population level (Supplementary Fig.~\num{10}), higher cell densities of $\Phi$F accumulated in the upstream reservoir compared to WT and $\Phi$A.
The same pattern was observed for all different channel widths. 
Thus, motility selection produced \ecoli phenotypes with graded rheotactic performance, with $\Phi$F showing enhanced upstream swimming in all cases.

Subsequently, we examined the structural basis of this rheotactic hierarchy. 
By labeling the bacterial flagella fluorescently (Methods), we found that $\Phi$F was hyperflagellated (Fig.~\ref{fig:2}F,G; Supplementary Figs.~11--13), with an average number of $N_f = 7\pm2$ flagella compared with $N_f = 4\pm1$ in wild-type cells. 
Conversely, $\Phi$A was hypoflagellated with only $N_f = 2\pm1$ flagella.
The flagella of $\Phi$F were also significantly longer, as measured along their contours (Supplementary Fig.~\num{11}A, B). Despite these differences in flagellar number and length, cell body dimensions, flagellar helical pitch and radius, intrinsic trajectory curvature, and the near-surface run-tumble frequency were indistinguishable among the three phenotypes (Supplementary Fig.~\num{11}C-G).

Next, we analyzed the genotypes of $\Phi$F and $\Phi$A. DNA sequencing revealed that $\Phi$F contained a large duplication of fourteen genes, comprising the majority of the \textit{flh} gene cluster (Supplementary Fig.~\num{14}). This included a duplication of the master regulator genes \textit{flhD} and \textit{flhC}. $\Phi$A contained an IS element insertion in \textit{rcsC}, which encodes a component of the Rcs atypical two-component regulatory system. This system regulates diverse cellular processes, including motility gene expression. These results directly link the enhanced upstream performance of $\Phi$F to increased flagellar abundance rather than to gross changes in body morphology or baseline swimming kinematics.

Because directed evolution can introduce additional genetic and physiological changes beyond flagellation, we next tested whether increasing flagellar abundance alone was sufficient to enhance rheotaxis (Fig.~\ref{fig:2}H). We genetically tuned flagellar abundance by inducibly expressing the flagellar master regulator \textit{flhD/C}~\cite{Lisevich2025-lw}. Indeed, increasing \textit{flhD/C} induction elevated flagellar abundance and enhanced upstream bias in rheotaxis assays, while producing negligible changes in intrinsic swimming speed under the same conditions (Supplementary Fig.~\num{15}). This targeted perturbation recapitulated the enhanced upstream swimming observed in the selected hyperflagellated phenotype, identifying flagellar abundance and length as key parameters controlling rheotaxis.

\subsection*{Cell curvature and type IV pili do not significantly affect rheotaxis}

To test whether non-flagellar traits could explain the reduced upstream performance of polar-flagellated species, we altered other cellular characteristics in \vcholerae (Supplementary Fig.~\num{16}). 
Deletion of \textit{crvA}, which straightens its curved cell body~\cite{Bartlett2017-vm,Martin2021-oc}, did not substantially alter rheotaxis, both on flat surfaces and in microstructured channels (Supplementary Fig.~\num{16}). 
Similarly, deletion of \textit{mshA}, which removes mannose-sensitive hemagglutinin pili~\cite{Utada2014-zu}, did not enhance upstream swimming and even reduced the upstream bias on flat surfaces (Supplementary Fig.~\num{16}E,F). These perturbations indicate that body curvature and pili are not the dominant factors that limit the upstream migration of \vcholerae under the tested shear conditions.

As another control, we asked whether rheotaxis extends to other peritrichous species beyond \ecoli (Supplementary Fig.~\num{17}). 
Indeed, \senterica also exhibited a pronounced upstream bias, comparable or even more than wild-type \ecoli. 
Specifically, both peritrichous species maintained a much larger upstream swimming velocity than the monotrichous species, for all shear rates tested, and a much larger fraction of cells moving against the flow. 
These results further emphasize the importance of multiflagellarity in rheotaxis.

\subsection*{Holographic 3D microscopy links flagellar arrangement to surface stability}

Because upstream swimming requires cells to remain near surfaces~\cite{Li2009-jh,Bianchi2017-eb,Mathijssen2019-zy}, we used digital in-line holographic microscopy (DIHM) \cite{Altman2021HolographicApproximation, Martin2022In-lineAnalysis} to measure the distance of individual cells from the surface. 
The bacterial holograms (Fig.~\ref{fig:3}A) were analyzed with Lorenz-Mie scattering theory \cite{Altman2020CATCH:Networks, Altman2023MachineTime} to reconstruct their 3D swimming trajectories (Fig.~\ref{fig:3}B; Methods).
Hence, we directly measured their average height $\langle z \rangle$ above the coverslip for different bacteria strains (Fig.~\ref{fig:3}C and Supplementary Fig.~\num{18}). 

Interestingly, the wild-type \ecoli and \senterica trajectories were observed to be concentrated in a narrow region near the surface, with $\langle z \rangle \sim \SI{1.6}{\micro\meter}$, where shear-surface coupling drives upstream alignment.
In contrast, \vcholerae and \paeruginosa exhibited more frequent excursions away from the boundary, leading to larger values of $\langle z \rangle \sim \SI{2.3}{\micro\meter}$ and $\SI{2.7}{\micro\meter}$, respectively. These excursions are expected to weaken upstream rheotaxis by reducing residence times near the surface and exposing cells to stronger advective flows, thereby limiting their ability to reorient and sustain upstream motion.
The same trend was observed when comparing wild-type \ecoli with the evolved strains from the motility selection experiments: The hyperflagellated strain ($\Phi$F) swam closer to the cover slip than the hypoflagellated strain ($\Phi$A), confirming that an increased number of flagella enhances surface accumulation.

These three-dimensional observations were consistent with measurements of the upstream swimming velocity  (Fig.~\ref{fig:3}D): This velocity can be decomposed into two contributions, $V_x = V_x^\textmd{swim} + V_x^\textmd{flow}$, from active swimming and passive advection, where the latter depends on the shear rate and the average height, $V_x^\textmd{flow} \approx - \dot{\gamma} \langle z \rangle$. Therefore, the slopes in Fig.~\ref{fig:3}D are a measure of the average height, $\partial V_x/\partial \dot{\gamma} \approx - \langle z \rangle$. 
Comparing these slopes across bacterial strains shows that \paeruginosa and \vcholerae (steeper slopes) swam significantly farther from the surface than \ecoli and \senterica (milder slopes), in agreement with the results from holography.

These height measurements provide a mechanistic bridge between morphology-dependent single-cell rheotaxis and population-scale invasion: Multiflagellated bacteria remained stably coupled to the boundary and reoriented better upstream, while uniflagellated cells more readily detached into the bulk flow, where downstream advection dominates.

\subsection*{Simulations reveal how multiflagellarity promotes upstream alignment}

To establish the biophysical mechanism that underpins how multiflagellarity enhances upstream swimming, we used low-Reynolds-number fluid--structure simulations that fully resolve the morphology and flagellar arrangement of motile bacteria: 
The cell body was represented as a rigid round-ended cylinder, and the flagella as elastic helical filaments that are coupled to a rotary motor by a flexible, straight hook (Fig.~\ref{fig:4}A, B; Supplementary Fig.~\num{19}). 
This model builds on our previous 
work \cite[see e.g.][]{Lee2021AFlagella,Park2019, Kim2022EffectsMotility,Lee2023,Fast2026SwimmingFlow}, including our treatment of shear flow near a no-slip surface, where wall-mediated hydrodynamic interactions were computed using the method of images for regularized Stokeslet formulation within the general immersed boundary framework (Methods). 
These simulations allowed us to independently vary the flagellar number and flagellar length, ceteris paribus, while tracking cell position, orientation, height above the boundary, and upstream migration.

We first compared monotrichous ($N_f=1$, orange) and peritrichous ($N_f=4$, green) swimmers initialized facing upstream under shear flow ($\dot{\gamma}=\SI{10}{\per\second}$). 
Monotrichous swimmers rapidly reoriented downstream and were advected with the flow, whereas peritrichous swimmers maintained stable upstream migration near the surface (Fig.~\ref{fig:4}C; Supplementary    Fig.~\num{20}). We then initialized both swimmer types at four different in-plane angles $\phi$ relative to the flow: $\phi=0^\circ, 90^\circ, 180^\circ$, and $270^\circ$. For all initial orientations, monotrichous swimmers were advected downstream and reoriented toward downstream-facing states. In contrast, peritrichous swimmers ultimately reoriented against the flow direction, with a slight bias to the right (positive vorticity direction), and stably migrated upstream (Fig.~\ref{fig:4}D; Supplementary Figs.~\num{21} and \num{22}).

Our simulations showed that multiflagellarity promotes upstream motility by enhancing reorientation against the flow and stabilizing near-surface residence. Varying the flagellar number revealed a shift in orientation: swimmers with fewer flagella tended to orient downstream, whereas those with more flagella preferentially aligned upstream (Fig.~\ref{fig:4}E; Supplementary Fig.~\num{24}). Swimmers with one or two flagella adopted positive pitch angles $\theta$, corresponding to nose-up orientations in which the cell body points away from the surface and is more readily detached. By contrast, swimmers with more than two flagella adopted negative pitch angles, corresponding to nose-down orientations toward the surface, stabilizing near-wall swimming (Fig.~\ref{fig:4}F; Supplementary Figs.~\num{23} and \num{24}). Simulations of four-flagellated swimmers with varying flagellar lengths further showed that longer flagella promoted stronger upstream alignment and more stable nose-down orientations near the surface (Fig.~\ref{fig:4}G, H; Supplementary Figs.~\num{25} and \num{26}). 

\section*{Discussion}

By integrating upstream swimming competition assays, single-cell tracking, directed evolution assays, 3D holography, and hydrodynamic simulations, we identify flagellar arrangement as a key determinant of bacterial upstream motility, and we uncover a unified mechanism: 
Multiflagellarity stabilizes near-wall residence and promotes upstream reorientation, enabling peritrichous bacteria to remain within the rheotactic zone and migrate upstream, whereas monotrichous swimmers more readily detach from the surface and are advected downstream.
Crucially, rheotaxis is governed not primarily by swimming speed, but by the coupling between propulsion, surface residence, and shear-driven reorientation.

These results suggest that the advantage of multiflagellarity may be particularly pronounced in confined flows relevant to host-associated ducts and biomedical devices, where sustained near-wall swimming can determine whether cells invade upstream or are washed downstream. In our narrow channels, the critical shear rate for upstream swimming, the highest shear rate at which cells can still swim against the flow, is $\dot{\gamma}_c \sim 80~\mathrm{s}^{-1}$, substantially higher than the $\sim 6.4~\mathrm{s}^{-1}$ reported for \textit{E. coli} on a planar surface~\cite{Kaya2012-po} and the $\sim 65~\mathrm{s}^{-1}$ reported in non-Newtonian polymer fluids~\cite{Cao2024-nm,Torres-Maldonado2024-mp}. Because $\dot{\gamma}_c$ sets an upper bound, upstream swimming remains possible in any flow whose shear rate falls below it: this includes urinary catheters ($1$--$5~\mathrm{s}^{-1}$), respiratory mucus flows ($5$--$20~\mathrm{s}^{-1}$), and cardiovascular currents ($20$--$120~\mathrm{s}^{-1}$)~\cite{Figueroa-Morales2020-mb,Ramirez-SanJuan2020Multi-scaleArrays,Martinez-Calvo2023ActiveEnvironments}. Moreover, the channel widths studied here span dimensions relevant to biological microchannels ($1$--$100~\mu\mathrm{m}$) and biomedical devices ($10$--$500~\mu\mathrm{m}$). Together, these comparisons suggest that multiflagellarity could promote upstream colonization under physiologically relevant conditions by enhancing near-wall residence and upstream reorientation, rather than by increasing swimming speed alone~\cite{Ramirez-SanJuan2020Multi-scaleArrays,Vasilev2009AntibacterialDevices,Cangui-Panchi2022Biofilm-formingReview}.

The mechanisms by which pathogenic bacteria navigate flow environments are key for clinical treatment of bacterial infections and understanding of pathogenesis. We recently showed that antibiotic treatment can affect upstream swimming via induction of cell elongation~\cite{Tao2026-jz}, and many bacteria—including some tested in this study—encounter flow during the course of human infection. Uropathogenic \ecoli (UPEC) is the number one cause of urinary tract infections (UTI) worldwide~\cite{Foxman2003-mn}. \ecoli is likely to encounter strong urine flow during UTI, and the excellent upstream swimming ability we have characterized may play a role in its robust uropathogenesis. \paeruginosa is a common cause of lung infections in the immunocompromised and is particularly deadly in cystic fibrosis (CF) patients~\cite{Malhotra2019-vt}, where it can be expected to encounter flow generated by lung cilia~\cite{Ramirez-SanJuan2020Multi-scaleArrays,Boeck2025-dm}. We demonstrated in this study that \vcholerae performs poorly when swimming against flow, particularly when considering its very high swimming velocity compared to the other bacterial species tested. Swimming against flow may be less important for \vcholerae, since it colonizes the small intestine after being ingested orally ~\cite{van-Kessel2024-ok}. This means it reaches its infection site by moving with flow, rather than fighting against it. We showed that the curved shape of the \vcholerae cell does not augment its ability to swim against flow, but shape has been shown to be key for pathogenesis by facilitating penetration of the intestinal mucus layer~\cite{Bartlett2017-vm}. These considerations suggest that the ability to navigate flow is key for determining individual bacterial species’ primary mechanisms of infection.

A related microbiological question is whether this second-scale rheotactic mechanism is coupled to slower regulatory responses. In many flow-exposed habitats, bacteria experience shear for hours to days, long enough to remodel motility, adhesion, and biofilm programs \cite{Conrad2018ConfinedBiofilms, Persat2015-rl, Padron2023BacteriaFlow}. Recent work shows that flagellar load, surface contact, and flow can regulate flagellar gene expression, stator recruitment, c-di-GMP signaling, and early biofilm development~\cite{Subramanian2019FunctionalFlagella, Dufrene2020Mechanomicrobiology:Forces, Laganenka2020-ht, Vrabioiu2022-dk, Hubert2024-hp, Wei2025-fr}. These findings suggest that bacteria may tune flagellar length or abundance in response to mechanical cues, thereby modulating upstream migration over longer timescales. Testing this hypothesis will require linking transcriptional reporters and mechanosensory perturbations to single-cell rheotaxis measurements under controlled shear conditions.

\small
\section*{Methods}

\subsection*{Bacterial strains and culture conditions}

The primary strain used in this study was \ecoli\ K-12 strain EPB47, an MG1655 derivative carrying a chromosomal \textit{ompA-cfp} fusion (a gift from Mark Goulian, University of Pennsylvania). Additional species included \paeruginosa\ PAO1, \vcholerae\ C6706, and \senterica\ serovar Typhimurium SL1344; all strains and plasmids are listed in Supplementary Table~1. Cells were revived from frozen glycerol stocks and streaked onto LB agar plates (1\% Bacto tryptone, 0.5\% yeast extract, 1.0\% NaCl, 1.5\% agar), followed by overnight incubation at \SI{32}{\degreeCelsius}. Single colonies were inoculated into \SI{3}{\milli\liter} LB medium and grown overnight at \SI{32}{\degreeCelsius} with shaking at \SI{250}{\revolutionsperminute}. 

Overnight cultures were diluted $10^{-2}$ into TB medium (1\% Bacto tryptone, 0.5\% NaCl) and grown to mid-log phase (OD$_{600}\approx 0.3$). For \ecoli, cells were subsequently transferred into Berg's motility buffer (BMB: \SI{0.1}{mM} EDTA, \SI{0.001}{mM} L-methionine, \SI{10}{mM} sodium lactate, \SI{67}{mM} NaCl, \SI{6.2}{mM} \ce{K2HPO4}, \SI{3.9}{mM} \ce{KH2PO4}), supplemented with L-serine and polyvinylpyrrolidone (PVP), and adjusted to pH 7.05. Cells were allowed to adapt in BMB for \SI{30}{\minute} prior to experiments. All microfluidic measurements were performed within \SI{2}{\hour} of preparation to ensure stable motility.

Directed motility selection and inducible \textit{flhD/C} modulation in \ecoli, as well as \textit{crvA} and \textit{mshA} deletion in \vcholerae, were performed as described in the Supplementary Information.

\subsection*{Fluorescent labeling of bacterial flagella}

Flagellar filaments were fluorescently labeled with an amine-reactive succinimidyl ester dye~\cite{Turner2000-fo}. Because free primary amines compete for the dye,
all labeling and washing steps were performed in a labeling buffer consisting of BMB prepared
without L-methionine or L-serine (\SI{0.1}{mM} EDTA, \SI{67}{mM} NaCl, \SI{6.2}{mM} \ce{K2HPO4},
\SI{3.9}{mM} \ce{KH2PO4}), supplemented with \SI{0.0001}{\percent} Tween-20 and adjusted to
pH~\num{7.05}. Mid-log-phase \ecoli cells were harvested by centrifugation at \SI{1000}{\times g}
for \SI{10}{\minute} at room temperature and gently washed twice in labeling buffer. The final
cell pellet was resuspended in approximately \SI{500}{\micro\liter} of labeling buffer, after
which \SI{25}{\micro\liter} of \SI{1}{M} sodium bicarbonate was added to raise the pH to
approximately \num{7.8}. AF 594 NHS ester (\SI{5}{\milli\gram\per\milli\liter} in DMSO) was then
added at \SI{10}{\micro\liter} per \SI{500}{\micro\liter} of cell suspension, and the suspension
was incubated for \SI{45}{\minute} at room temperature in the dark under gentle gyration at
approximately \SI{100}{\revolutionsperminute}. Cells were subsequently washed three to four times
in labeling buffer to remove unreacted dye and gently resuspended for fluorescence imaging. This
amine-specific labeling approach enabled visualization of intact flagellar filaments while
retaining bacterial motility.

\subsection*{Genetic analysis and mutant construction}

To identify mutations conferring increased or decreased swim ring diameter and flagellar number, $\Phi$A and $\Phi$F were analyzed via whole-genome sequencing. Genomic DNA was extracted from stationary-phase cultures via phenol/chloroform extraction and was sequenced using long-read nanopore sequencing (Plasmidsaurus).

To achieve tunable flagellar number and motility gene expression, the genes \textit{flhD} and \textit{flhC} were PCR-amplified from wild-type \textit{E. coli} genomic DNA and cloned into pBBR1-MCS2 following standard restriction enzyme cloning procedures. The resulting construct was verified via whole-plasmid sequencing (Plasmidsaurus), and the plasmid was introduced into an \textit{E. coli} EPB47 strain containing a chromosomal \textit{flhDC} inactivation via chemical transformation. Flagellar gene expression was induced via addition of IPTG. Cells retained motility in the absence of inducer, likely due to plasmid copy number and leaky expression from the \textit{lac} promoter.

\subsection*{Microfluidic device fabrication and flow control}

Microfluidic devices were fabricated in polydimethylsiloxane (PDMS) using standard soft lithography from SU-8 masters. PDMS and curing agent were mixed at a 10:1 (w/w) ratio, degassed, cast onto the master, and cured overnight at \SI{65}{\degreeCelsius}. Devices were bonded to glass coverslips following oxygen plasma treatment and thermally reinforced at \SI{95}{\degreeCelsius}.

Structured channels had depth $H=\SI{10}{\micro\meter}$ and widths $W$ ranging from \SI{10}{\micro\meter} to \SI{50}{\micro\meter}. Flat-surface assays were performed in wide channels ($H=\SI{50}{\micro\meter}$, $W=\SI{500}{\micro\meter}$) to minimize side-wall effects.

Flow was imposed using a pressure-driven controller (ElveFlow OB1 MK4), generating steady Poiseuille flow. Shear rates were defined either as near-wall shear (flat surfaces) or as edge-averaged shear (structured channels), as detailed in the Supplementary Information.

\subsection*{Imaging and trajectory analysis}

Bacterial dynamics were recorded on a Nikon TI2-E microscope using a \num{20}$\times$ objective (NA 0.75) and an sCMOS camera (Hamamatsu Orca-Fusion Gen-III) up to 100 FPS. Bright-field imaging was used for single-cell tracking, while fluorescence imaging was used for population-level redistribution assays.

For structured-channel and invasion experiments, imaging was performed at the channel mid-plane. For flat-surface assays, the focal plane was positioned near the bottom boundary to capture cells within approximately \SI{5}{\micro\meter} of the surface.

Videos were processed in ImageJ, and trajectories were extracted using TrackMate with custom batch macros. Only trajectories longer than \SI{1}{\second} were included in quantitative analyses. Streamwise velocities, curvature, reorientation frequency, and upstream fractions were computed using custom Python scripts. Three-dimensional height distributions were obtained using digital holographic microscopy as described in the Supplementary Information.

\subsection*{Holographic analysis}

Three-dimensional bacterial positions were measured using an in-line holographic imaging setup built on a Nikon TI2-E microscope equipped with a \SI{450}{\nano\meter} fiber-coupled laser diode. The laser beam was collimated and aligned approximately normal to the imaging plane, and holograms were recorded with an sCMOS camera through a 40$\times$ objective at \SIrange{10}{30}{fps}. Raw holograms were corrected using a running normalization procedure before fitting~\cite{Altman2021HolographicApproximation, Martin2022In-lineAnalysis}.

Axial positions were extracted by fitting the recorded holographic intensity patterns to Lorenz--Mie scattering theory using the \texttt{pylorenzmie} package~\cite{Altman2020CATCH:Networks, Altman2023MachineTime}. In this approach, each bacterium was approximated as an effective scattering sphere, yielding best-fit estimates of the lateral position, axial position, effective diameter and refractive index. Although bacterial cells are elongated rather than spherical, the effective-sphere approximation provides a robust estimate of axial position for tracking near-wall height distributions. The resulting three-dimensional trajectories were used to quantify bacterial height above the surface and compare near-wall residence across species.

\subsection*{Modeling and simulations}

We model a bacterium as a rod-shaped cell body connected to one or more helical flagella through a short, compliant hook. The cell body dynamics is represented using a penalty-based rigid-body formulation, while the elastic flagellum is described by the unconstrained Kirchhoff rod theory~\cite{Lim2008, Olson2013, Lee2021AFlagella}. Hydrodynamic interactions between the bacterium and the surrounding viscous fluid are resolved using the regularized Stokeslet formulation together with the method of images to account for hydrodynamic interactions with a nearby planar surface~\cite{Cortez2001,Kim2022EffectsMotility, Cortez2005, Ainley2008, Park2019, Lee2023}.

The bacterial flagellar motor generates a constant torque to rotate the flagellum, while an equal and opposite reaction torque acting on the cell body causes it to counterrotate. The swimming motion of the bacterium is determined by enforcing total force-free and torque-free conditions. This modeling framework captures both the direct mechanical coupling and the hydrodynamic interactions among the rigid cell body, the motor-driven flexible flagella, and the planar surface. A detailed mathematical formulation is provided in the Supplementary Material.

\normalsize
\section*{Acknowledgments}

We thank all members of the Mathijssen lab for their support and insightful discussions. R.T. was supported by Dissertation Completion Fellowship from the University of Pennsylvania. W.L. was supported in part by a National Institute for Mathematical Sciences (NIMS) grant funded by the Korean government (B26910000), and in part by a National Research Foundation of Korea (NRF) grant funded by the Korean government (MSIT) (RS-2025-16071334). Y.K. was supported by National Research Foundation of Korea Grant funded by the Korean Government (RS2026-25478035). J.Z. was supported by the National Institute of Allergy and Infectious Diseases of the National Institutes of Health (R01AI178908 and R01AI157106). S.L. was supported by NSF (CBET-2415406) and the Charles Phelps Taft Research Center at the University of Cincinnati. A.J.T.M.M. acknowledges funding from the Charles E. Kaufman Foundation (Early Investigator Research Award KA2022-129523; and New Initiative Research Award KA2024-144001), the National Science Foundation (Career Award CBET-2542731; and UPenn MRSEC DMR-2309043), the University of Pennsylvania (URF, CURF, VIPER, Vagelos MLS, and FERBS programs), and the Research Corporation for Science Advancement (Cottrell Scholar Award CS-CSA-2026-125). This work was carried out in part at the Singh Center for Nanotechnology, which is supported by the NSF National Nanotechnology Coordinated Infrastructure Program under grant NNCI-2025608.

\section*{Data availability}

All datasets generated and analysed during the current study are
available from the corresponding authors on request. Source data are provided with
this paper.

\section*{Author contributions}
R.T., N.C.E., and W.L. contributed equally to this work. R.T. and A.J.T.M.M. designed the
research. R.T. designed and performed the microfluidic and single-cell tracking
experiments, the directed evolution assays and the fluorescent flagellar labeling, and
analyzed the resulting data. N.C.E. performed the genetic construction and the
whole-genome sequencing analysis. W.L., Y.K. and S.L. developed and performed the
fluid--structure simulations. L.A. and R.T. performed the holographic imaging and
analyzed the holographic data. D.G. assisted with the directed evolution experiments
and the fluorescent flagellar labeling. L.C. and L.L. performed the scanning electron
microscopy imaging. A.J.T.M.M., S.L. and J.Z. supervised the study. R.T., A.J.T.M.M.
and S.L. wrote the manuscript with input from all authors.

\section*{Competing interests}
The authors declare no competing interests.

\printbibliography[title={References}]

\clearpage
\input{Figures}

\end{document}

%% file: Figures.tex

\begin{figure}
    \centering
    \includegraphics[width=0.95\textwidth]{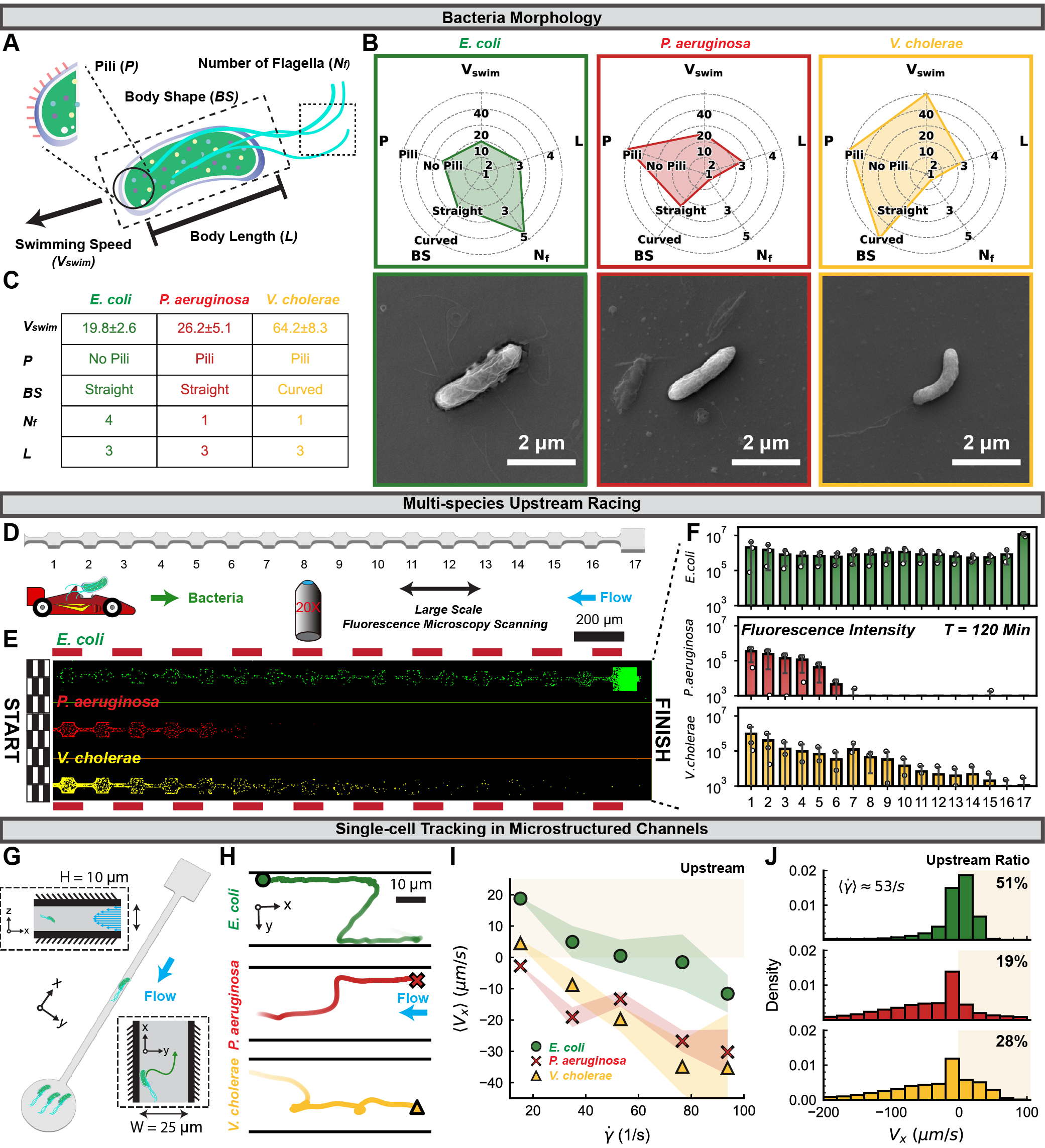}
    \caption{
    \textbf{Upstream swimming competitions comparing bacterial species.}
    (\textbf{A-C}) Juxtaposition of morphological properties and flagellar arrangements of \textit{E.~coli}, \textit{P.~aeruginosa}, and \textit{V.~cholerae}, with representative SEM images. Swimming speeds in (\textbf{C}) are the mean of $N = 3$ independent experiments $\pm$ the standard deviation across those three experiment-level means ($N_{\mathrm{traj}} > 500$).
    (\textbf{D}) Schematic of population-scale upstream racing assay, showing the microfluidic channel geometry and imposed flow.
    (\textbf{E}) Large-scale scanning fluorescence microscopy reveals how the different species of bacteria move upstream (from left to right), under identical shear conditions ($\dot{\gamma} = \SI{20}{\per\second}$).
    (\textbf{F}) Quantification of fluorescence intensity, proportional to number of bacteria, in upstream reservoir (\#17) after \SI{120}{\minute}. Bars show the mean of $N = 3$ independent experiments, open circles the individual replicates, and error bars the standard deviation across replicates.
    (\textbf{G}) Schematic of individual-cell-scale upstream racing assay in a thin microfluidic channel.
    (\textbf{H}) Representative single-cell trajectories.
    (\textbf{I}) Ensemble-averaged upstream swimming velocities $\langle V_x \rangle$ as a function of shear rate. Each point uses $N_{\mathrm{traj}} > 500$ trajectories; symbols show the ensemble mean, and shaded bands show the standard deviation across $N = 3$ independent experiments at each shear rate.
    (\textbf{J}) Distributions of this upstream swimming velocity, and fraction of bacteria moving upstream, for each species.
    }
    \label{fig:1}
\end{figure}


\begin{figure}
    \centering
    \includegraphics[width=0.95\textwidth]{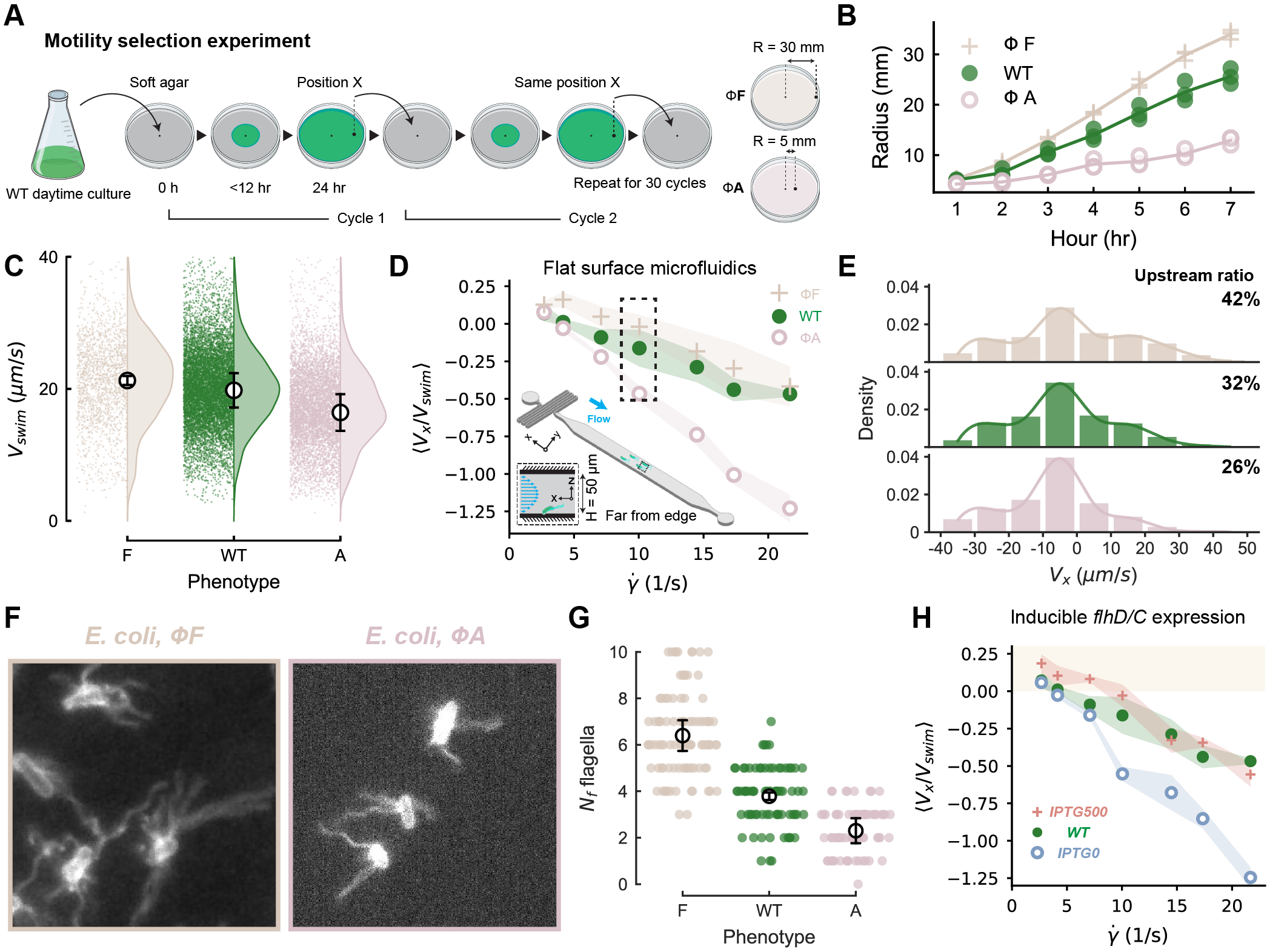}
    \caption{
    \textbf{Directed evolution produces hyperflagellated strains that are hyperrheotactic.}
    (\textbf{A}) Motility selection procedure used to isolate hypo- and hypermotile phenotypes. Repeated migration--isolation cycles on soft agar generate evolved populations (Phenotypes~F, $\Phi$F, and~A, $\Phi$A).
    (\textbf{B}) Expansion-speed measurements in soft agar for $\Phi$F, WT, and $\Phi$A. Points show the mean expansion radius and error bars the standard deviation across $N = 3$ independent replicates per time point.
    (\textbf{C}) Swimming-speed measurements of $\Phi$F, WT, and $\Phi$A cells.
    (\textbf{D}) Streamwise velocity normalized by swimming speed, $\langle V_x / V_{\mathrm{swim}} \rangle$, as a function of shear rate $\dot{\gamma}$ for $\Phi$F, WT, and $\Phi$A cells in flat-surface microfluidic assays. Each data point includes $N_{\mathrm{traj}} > 1000$ trajectories. Symbols denote the ensemble mean and shaded bands the standard deviation across $N = 3$ independent experiments at each shear rate.
    (\textbf{E}) Streamwise-velocity distributions for $\Phi$F, WT, and $\Phi$A cells. Percentages denote the fraction of upstream-swimming trajectories.
    (\textbf{F}) Representative fluorescence images of labeled flagella in $\Phi$F (left) and $\Phi$A (right) cells.
    (\textbf{G}) Flagellar number per cell for $\Phi$F, WT, and $\Phi$A. In (\textbf{C}) and (\textbf{G}), colored points represent individual measurements, white circles indicate the mean, and error bars denote $\pm 1$ s.d.; data were obtained from $N = 3$ independent experiments.
    (\textbf{H}) Mean normalized streamwise velocity as a function of shear rate for strains with induced \textit{flhD/C} expression (IPTG500), WT, and uninduced \textit{flhD/C} expression (IPTG0). Each data point includes $N_{\mathrm{traj}} > 1000$ trajectories. Symbols denote the ensemble mean and shaded bands the standard deviation across $N = 3$ independent experiments at each shear rate.
    }
    \label{fig:2}
\end{figure}


\begin{figure}
    \centering
    \includegraphics[width=0.95\textwidth]{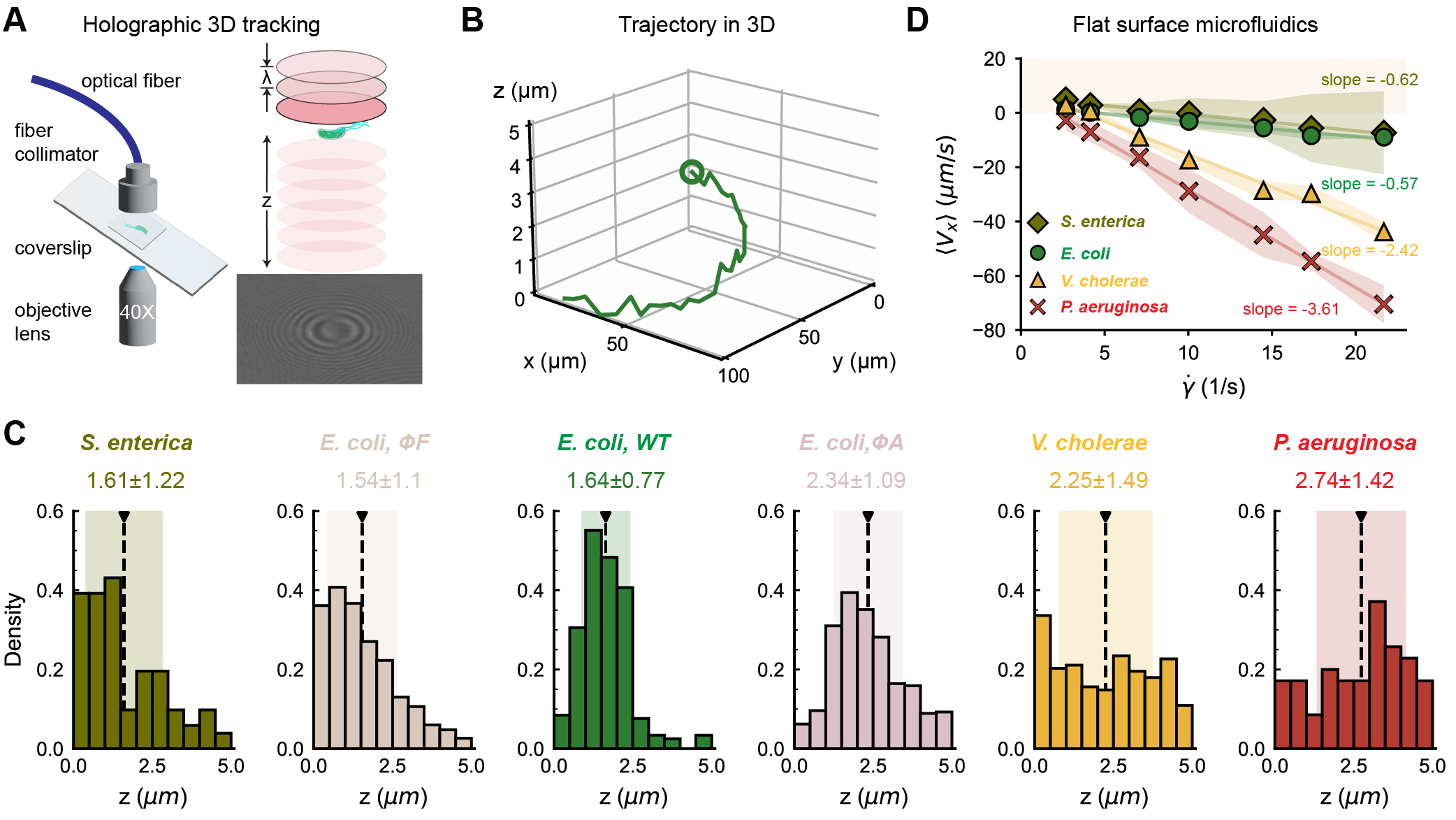}
    \caption{
    \textbf{Holography shows flagella stabilize near-surface motility.}
    (\textbf{A}) Holographic 3D tracking setup used to reconstruct bacterial trajectories and measure swimming height above a flat surface.
    (\textbf{B}) Representative reconstructed three-dimensional trajectory of \ecoli{} approaching and swimming near the surface.
    (\textbf{C}) Probability density distributions of swimming height $z$ within \SI{5}{\micro\meter} of the surface for \textit{S.~enterica}, \ecoli{} phenotype~F ($\Phi$F), wild-type \ecoli{}, \ecoli{} phenotype~A ($\Phi$A), \textit{V.~cholerae}, and \textit{P.~aeruginosa}. The numbers of trajectories analyzed were $N_{\mathrm{traj}} = 5$, 33, 23, 25, 21, and 8, respectively. In each distribution, the black dashed line and downward-pointing triangle indicate the mean height, and the shaded region represents one standard deviation about the mean.
    (\textbf{D}) Ensemble-averaged streamwise velocity $\langle V_x\rangle$ as a function of shear rate $\dot{\gamma}$ for \textit{S.~enterica} (olive), \ecoli{} (green), \textit{V.~cholerae} (yellow), and \textit{P.~aeruginosa} (red). Solid lines indicate unconstrained linear fits. Each data point includes $N_{\mathrm{traj}} > 1000$ trajectories. Symbols denote the ensemble mean and shaded bands the standard deviation across $N = 3$ independent experiments at each shear rate.
    }
    \label{fig:3}
\end{figure}


\begin{figure}
    \centering
    \includegraphics[width=0.95\textwidth]{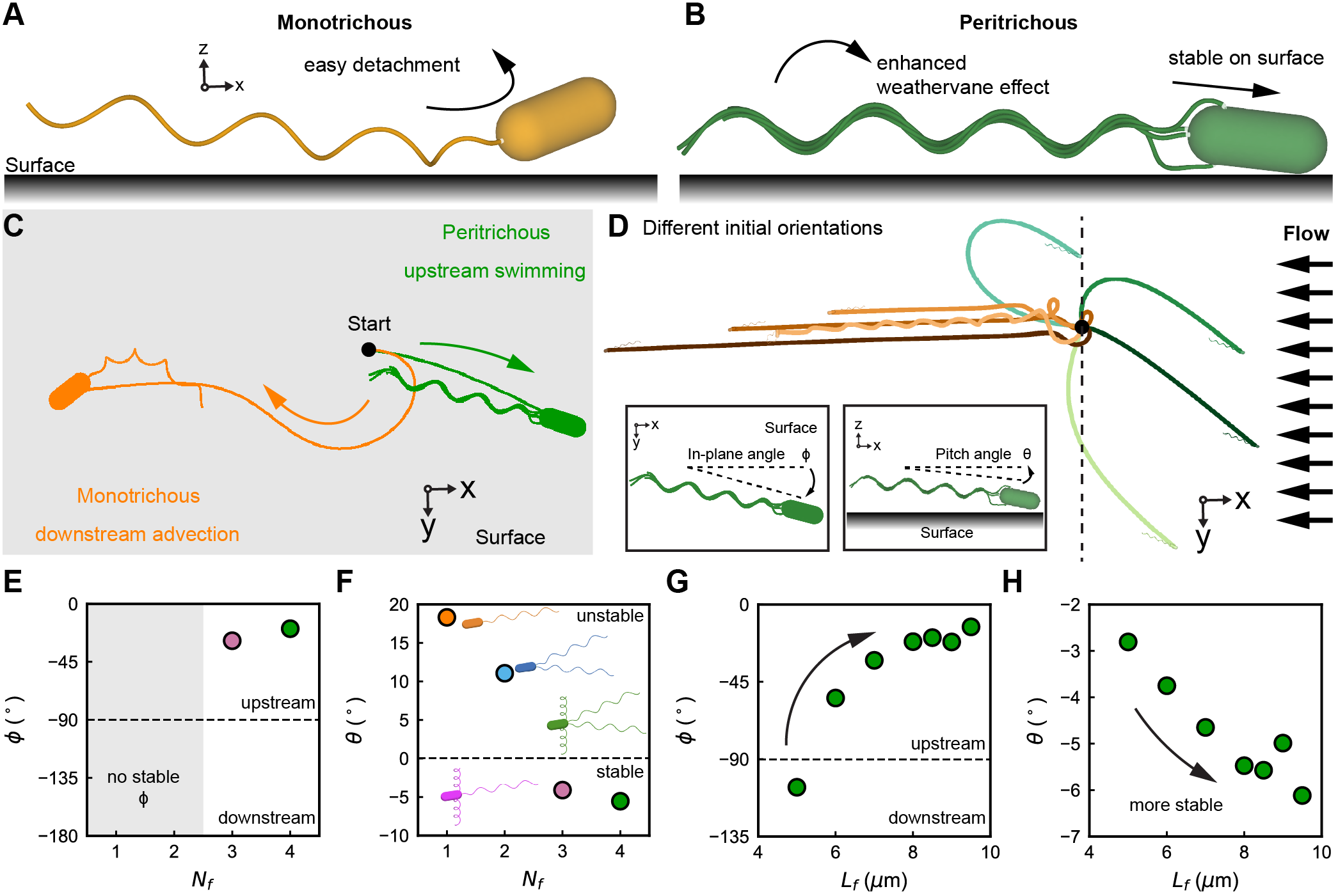}
    \caption{
    \textbf{Hydrodynamic simulations reveal how multiflagellarity promotes upstream alignment.}
    (\textbf{A}, \textbf{B}) Schematics of monotrichous and peritrichous swimmer configurations near a planar solid surface under shear flow.
    Simulations in (\textbf{C}--\textbf{E}) and (\textbf{G}) were performed at an imposed shear rate of $\dot{\gamma} = \SI{10}{\per\second}$, whereas simulations in (\textbf{F}) and (\textbf{H}) were performed without imposed shear flow.
    (\textbf{C}) Representative simulated trajectories of monotrichous (orange) and peritrichous (green) swimmers initialized near the same position and oriented against the imposed flow.
    (\textbf{D}) Representative simulated trajectories of monotrichous (orange) and peritrichous (green) swimmers initialized with different in-plane orientations relative to the imposed flow. Insets define the in-plane angle $\phi$ and pitch angle $\theta$.
    (\textbf{E}) Steady-state in-plane angle $\phi$ under shear flow as a function of flagellar number $N_f$. Values of $\phi$ above $\SI{-90}{\degree}$ denote upstream-oriented states, whereas values below $\SI{-90}{\degree}$ denote downstream-oriented states.
    (\textbf{F}) Pitch angle $\theta$ in simulations without imposed shear flow as a function of flagellar number $N_f$. Negative values of $\theta$ denote nose-down orientations toward the surface.
    (\textbf{G}) Steady-state in-plane angle $\phi$ under shear flow as a function of flagellar length $L_f$ for swimmers with four flagella.
    (\textbf{H}) Pitch angle $\theta$ in simulations without imposed shear flow as a function of flagellar length $L_f$ for swimmers with four flagella.
    }
    \label{fig:4}
\end{figure}
